\documentclass[twocolumn]{aastex62}

\usepackage{verbatim, graphicx,  ifthen, pbox}
\usepackage{eso-pic}
\usepackage{CJKutf8}
\usepackage{xcolor}
\usepackage{hyperref}
\usepackage{breakurl}
\usepackage{amsmath}
\usepackage{graphicx}
\usepackage{verbatim}
\usepackage{booktabs}
\usepackage[T1]{fontenc}
\usepackage{ae,aecompl}
\usepackage{statmath}

\usepackage{newtxtext,newtxmath}

\begin{document}
\title{Generating Stellar Obliquity in Systems with Broken Protoplanetary Disks}
\shorttitle{Broken Disks \& Stellar Obliquity}
\shortauthors{Epstein-Martin, Becker \& Batygin}


\author[0000-0001-9310-7808]{Marguerite Epstein-Martin}
\affiliation{Division of Geological and Planetary Sciences, California Institute of Technology, Pasadena, CA 91125, USA}
\affiliation{Department of Astronomy, Columbia University, 550 West 120th Street, New York, NY 10027, USA}
\author[0000-0002-7733-4522]{Juliette Becker}
\affiliation{Division of Geological and Planetary Sciences, California Institute of Technology, Pasadena, CA 91125, USA}
\author[0000-0002-7094-7908]{Konstantin Batygin}
\affiliation{Division of Geological and Planetary Sciences, California Institute of Technology, Pasadena, CA 91125, USA}
\correspondingauthor{Marguerite Epstein-Martin}
\email{m.epstein-martin@columbia.edu}

\accepted{March 4, 2022}

\begin{abstract}
Recent advances in sub-millimeter observations of young circumstellar nebulae have opened an unprecedented window into the structure of protoplanetary disks, which has revealed the surprising ubiquity of broken and misaligned disks. In this work, we demonstrate that such disks are capable of torquing the spin axis of their host star, representing a hitherto unexplored pathway by which stellar obliquities may be generated. The basis of this mechanism is a crossing of the stellar spin precession and inner disk regression frequencies, resulting in adiabatic excitation of the stellar obliquity. We derive analytical expressions for the characteristic frequencies of the inner disk and star as a function of the disk gap boundaries, and place an approximate limit on the disk architectures for which frequency crossing and resulting obliquity excitation are expected, thereby illustrating the efficacy of this model. Cumulatively, our results support the emerging concensus that significant spin-orbit misalignments are an expected outcome of planet formation.

\end{abstract}

\section{Introduction} \label{sec:intro}
The standard picture of planet and star formation holds that planetary systems are born in a disk, yielding an initially common plane. This theory is validated by the near-alignment of our own Solar System and numerous multi-transiting exoplanet systems and has remained largely unaltered since its conception by Kant and Laplace in the 18th Century \citep{Kant1755, Laplace1796}. A natural logical extension is to assume that coincidence of angular momentum vectors extends to stellar spin-axes.

The soaring number of exoplanet discoveries over the past two decades \citep{Borucki2010, Borucki2011, Howell2014, Guerrero2021} and the wide diversity of systems represented, has, however, required a re-evaluation of the ‘one-size-fits-all’ paradigm. In particular, measurement of misalignment between stellar spin axis and planetary orbit, otherwise known as stellar obliquity, has increasingly fallen within the reach of the Rossiter-McLaughlin effect \citep{Rossiter1924, McLaughlin1924, Winn2005} as well as other techniques (see review by \citealt{Winn&Frabcycky2015}). Current measurements suggest that obliquities are common, particularly in systems hosting short-period planets orbiting stars hotter than the Kraft break \citep{Winn2010, Morton2014}.  

The dominant source of stellar obliquities remains controversial. Post-formation perturbations due to planet-planet scattering \citep{ Ford&Rasio2008, Nagasawa2008, Beauge&Nesvorny2012}, Kozai-Lidov cycling \citep{ Wu&Murray2003, Fabrycky&Tremaine2007, Naoz2011}, and chaotic interactions with additional planets in the system \citep{Lithwick&Wu2012} have all been proposed as plausible misalignment pathways. Alternative explanations, disrupting the established disk-star alignment paradigm, assert that obliquities may be excited prior to or in concert with planet formation. For example, non-axisymmetric collapse of the molecular cloud core has been shown to create stellar obliquities \citep{Tremaine1991, BateLodato&Pringle2010} as has interactions between the stellar magnetosphere and disk \citep{Lai2011}.  

Of particular relevance to this work is a distinct process: the binary-disk torquing mechanism for exciting spin-orbit misalignments \citep{Batygin2012,2013ApJ...778..169B,Spalding&Batygin2014, Lai2018}. Within the framework of this model, spin-orbit misalignments are induced by a stellar companion, which torques the nebula, resulting in nodal regression. Simultaneously, the nebula torques the star, driving nodal regression of the stellar spin-axis. As the nebula dissipates and the primary star contracts, a commensurability between the regressing disk and precessing stellar spin axis ensues, and the system evolves through a resonant encounter. This encounter leads to impulsive excitation of the stellar obliquity, replicating the full obliquity range. Operation of this mechanism is best exemplified by the recent discovery of K2-290 A, a star whose spin is tilted 124 degrees with respect to the orbits of both of its known planets, with a binary companion capable of tilting the protoplanetary disk \citep{Hjorth2021}.

Previous work towards understanding stellar obliquity with respect to short-period planets has generally treated disks as rigid objects, where rigidity means equal inclinations and precession of the nodes for all disk annuli. However, recent observations of protoplanetary disks show evidence for disks containing wide gaps with misalignments between the inner and outer components. To this end, \citet{Brauer2019} observed shadows cast by misalignment between inner and outer disks; \citet{Ansdell2020} observed `dipper' transit systems and found that often, the inner disk causing the dipper appears to be misaligned with an outer component visible with ALMA; \citet{Francis2020} used images of inner and outer disks to study the brightness profiles and identified additional misaligned systems. In fact, of all systems where both inner and outer disk components can be resolved by ALMA \citep{Francis2020}, roughly 85\% of protoplanetary disks exhibit some misalignment or indirect evidence of misalignment (such as warps, shadows, or a dipper host). 
Such misalignments can be caused by massive planets \citep{Nealon2018, Zhu2019} forming in the disk or by a stellar binary companion \citep{Facchini2018}.

While presenting a complicating detail to the previously mentioned explanations of stellar obliquity, the dynamical evolution of a broken, misaligned disk can be analyzed in much the same way as the binary-disk torquing scenario. That is, the outer, misaligned disk is qualitatively analogous to a binary companion, acting as a time-dependent outer potential influencing the inner disk-star orientation. Accordingly, in this work, we investigate the possibility of disk-disk torquing producing stellar obliquities independently, without invocation of a binary companion. 

We ground our analysis in the framework of secular theory, using angular momentum arguments in Section (\ref{sec:l_hierarchy}) to support a simplified analytical model, demonstrating that back-reactions of the star upon the inner disk and the inner disk upon the outer disk can be neglected. In Section (\ref{sec:grav}) we derive the characteristic precession timescales for the inner disk and stellar spin axis. We use these expressions in Section (\ref{sec:obliq}) to define disk structure criteria for obliquity excitation and demonstrate the efficacy of this mechanism with numerical integrations. In Section (\ref{sec:conclude}), we conclude and note avenues for future study. 

\begin{figure*}
\centering
 \includegraphics[width=5.5in]{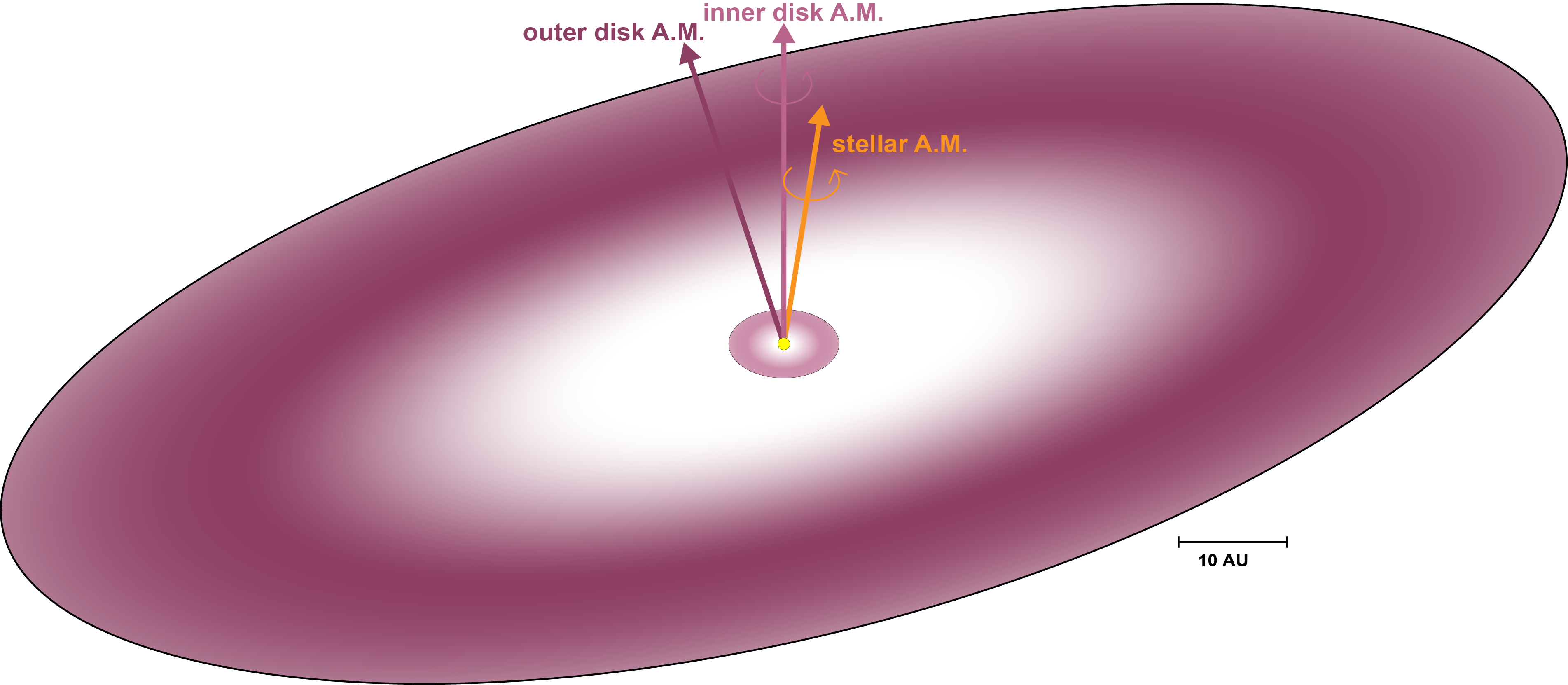}
 \caption{A schematic of the system geometry under consideration in this work, which include three distinct components: a central star, a inner disk, and an outer disk. Misalignment in angular momenta may exist between any components in the system, and the precession of the star is driven by the effectively static potential of the outer disk and the time-varying potential of the inner disk.}
    \label{fig:schematic}
\end{figure*}


\section{Angular Momentum Distribution in the Star-Disk-Disk System}
\label{sec:l_hierarchy}

To determine the relevant dynamics, we will first consider the hierarchy of angular momentum in the star-broken disk system. By comparing the angular momentum for the star, inner disk, and outer disk, we can determined which component drives which, and demonstrate that back-reactions can be neglected in this analysis. We note that this angular momentum argument holds provided that the effects of bending waves, viscosity, and disk self-gravity are sufficiently strong that the successive radial components of the disk remain coupled to each other \citep{Batygin2012, Lai2014, Zanazzi2018}. In this case, the individual disk components may be treated as effectively rigid, an assumption that is expected for typical protoplanetary disk parameters \citep{1996MNRAS.282..597L, 2011A&A...533A...7B, 2018MNRAS.477.5207Z}. 

As is typically done in the literature, for the protoplanetary disk, we will assume a surface density function of the type:
\begin{equation}
    \Sigma = \Sigma_{0}(a_{0} / a)^{p}\,.
\label{eq:surface_density}
\end{equation}
In Equation (\ref{eq:surface_density}), $a_{0}$  is a reference radius, $\Sigma_{0}$ is the surface density at radius $a_{0}$, and $p$ is the power law index of the surface density profile. For definitiveness, we assume $p = 1$ \citep{Mestel1963, Andrews2009}. Although the disk considered in this work has a gap separating the misaligned inner and outer components, we assume that when the disk is present, it obeys the above surface density profile.

To find the angular momenta of the two disk components, we first specify the angular momentum of a infinitesimal annulus,
\begin{equation}
    d\mathcal{L} = \sqrt{G M a} dm \,,
\end{equation}
where $G$ is the gravitational constant, $M$ the mass of the central star, $dm = 2 \pi \Sigma a da$ is the mass enclosed by the ring of width $da$ and mean orbital distance $a$. 
With this identity in mind, we take the disk to be made up of a series of infinitesimal annuli such that the expression for the angular momentum of the disk component can be written:
\begin{equation}
    \mathcal{L} = \int^{a_{\rm{outer}}}_{a_{\rm{inner}}} 2 \pi\ \sqrt{G M a}\ \Sigma_{0}\ (a_{0}/a)\ a da\,, \label{eq:disk}
\end{equation}
where $a_{\rm{inner}}$ and $a_{\rm{outer}}$ denote the inner and outer radii of the disk component under consideration. Similarly, we can write an expression for the angular momentum of the central star:
\begin{equation}
    \mathcal{L}_{*} = I_{0} M R^{2} \Omega \label{eq:star}
\end{equation}
where $I_{0}$ denotes the numerical coefficient of the stellar moment of inertia, $R$ the stellar radius, and $\Omega$ the stellar rotational frequency.

Naturally, a young, disk-bearing system can be expected to undergo a physical transformation over the course of the disk and stellar PMS lifetime. Thus, we cannot necessarily take the comparative relationships between the stellar, inner disk, and outer disk angular momenta (i.e. $    \mathcal{L}_{\rm{d,out}} > \mathcal{L}_{\rm{d,in}} > \mathcal{L}_{*} $) as a given. It is thus worthwhile to examine the time dependence of the components of angular momenta over the course of the system evolution.

The T Tauri stars with which we are concerned contract on a Kelvin-Helmholtz timescale as they proceed along their Hayashi track. Using a polytropic model to specify the stellar properties, the stellar radius is expected to evolve as \citep{Chandrasekhar1939}: 
\begin{equation}
R_{\star} = R_{\star}^0 \left[  1 + \left(\frac{5 - n }{3} \right)\frac{24 \pi \sigma T_{\rm{eff}}^4}{G M \left(R_{\star}^0\right)^3} t\right]\,.
\label{eq:stellar_radius}
\end{equation}
For a fully convective star, emerging from the embedded phase and evolving along the Hayashi track, a polytropic index $n = 3/2$ is appropriate.  $R_{\star}^0$ is the stellar radius at an initial time $t = 0$, approximately .1 to .5 Myr since the stellar birth. A star of mass $M = 1 M_{\sun}$ is expected to have an initial radius $R_{\star}^0 \approx 4 R_{\sun}$, evolving according to Equation (\ref{eq:stellar_radius}) with effective temperature $T_{\rm{eff}} = 4100$ K \citep{Seiss2000}.

The other physically evolving parameter we must consider here is the mass of the inner and outer disk, which will decrease with time. Here we follow \citet{Laughlin2004} and use a simple dissipation model for both disk components, varying mass with time as: 

\begin{equation}
M_{\rm{disk}} = \frac{M_{\rm{disk, 0}}}{1 + t/\tau_{\rm{acc}}} \,.
\end{equation}
A reasonable fit to observations, consistent with the work of \citet{Hillenbrand2008, Herczeg2008, Hartmann2008}, is achieved by setting the accretion timescale to $\tau_{\rm{acc}} = 0.5$ Myr and the initial surface density profile of the disk by Equation (\ref{eq:surface_density}) with $a_{0} = 1$ AU, and $\Sigma_{0} = 2 \times 10^3$ g cm$^{-2}$.  

To complete our calculation, we must define the radial bounds over which we integrate for both the inner and outer disks. The interior boundary of the inner disk is determined self-consistently, by noting that T Tauri magnetic fields of $\sim kG$ are expected to disrupt the conductive nebular fluid near the star, carving out a cavity in the inner disk. The radial extent of this disruption can be derived in several ways, all approximately yielding the same scaling, but most straightforwardly understood by equating the stellar magnetic pressure to the ram pressure of infalling disk material \citep{Ghosh&Lamb1978, Armitage2020}. The result determines the interior disk edge also known as the Alfvén radius: 

\begin{equation}
    a_{\rm{x}} = \Pi \left(\frac{8 \pi^2}{\mu_0^2} \frac{\mathcal{M}^4}{G M\dot{M_{\rm{d}}^2}} \right)^{1/7} \,.
\end{equation}
Here $\Pi$ is a dimensionless constant of order unity, $\mathcal{M}$ is the stellar magnetic moment, and $\dot{M_{\rm{d}}}$ is the mass accretion rate onto the central star. Because higher order components of the stellar magnetic field drop off steeply with radius, the magnetic moment is well approximated by a dipolar field ($\mathcal{M} = B_{\rm{dip}} R_{\star}^3/2$). For typical T Tauri parameters, the theoretically expected location of the interior disk edge $ a_{\rm{x}}$ will range between 0.05 - 0.2 AU \citep{Shu1994}. 

Assuming the outer disk extends to $\sim 100$ AU, we are left to define only the radial extent of the disk gap. We take these from observations by \citet{Ansdell2020, Francis2020} which find that on average the interior gap edge $A_{\rm{in}} \sim 5$ AU and the outer gap edge $A_{\rm{out}} \sim 50$ AU.  

\begin{figure}
\centering
 \includegraphics[width=3.35 in]{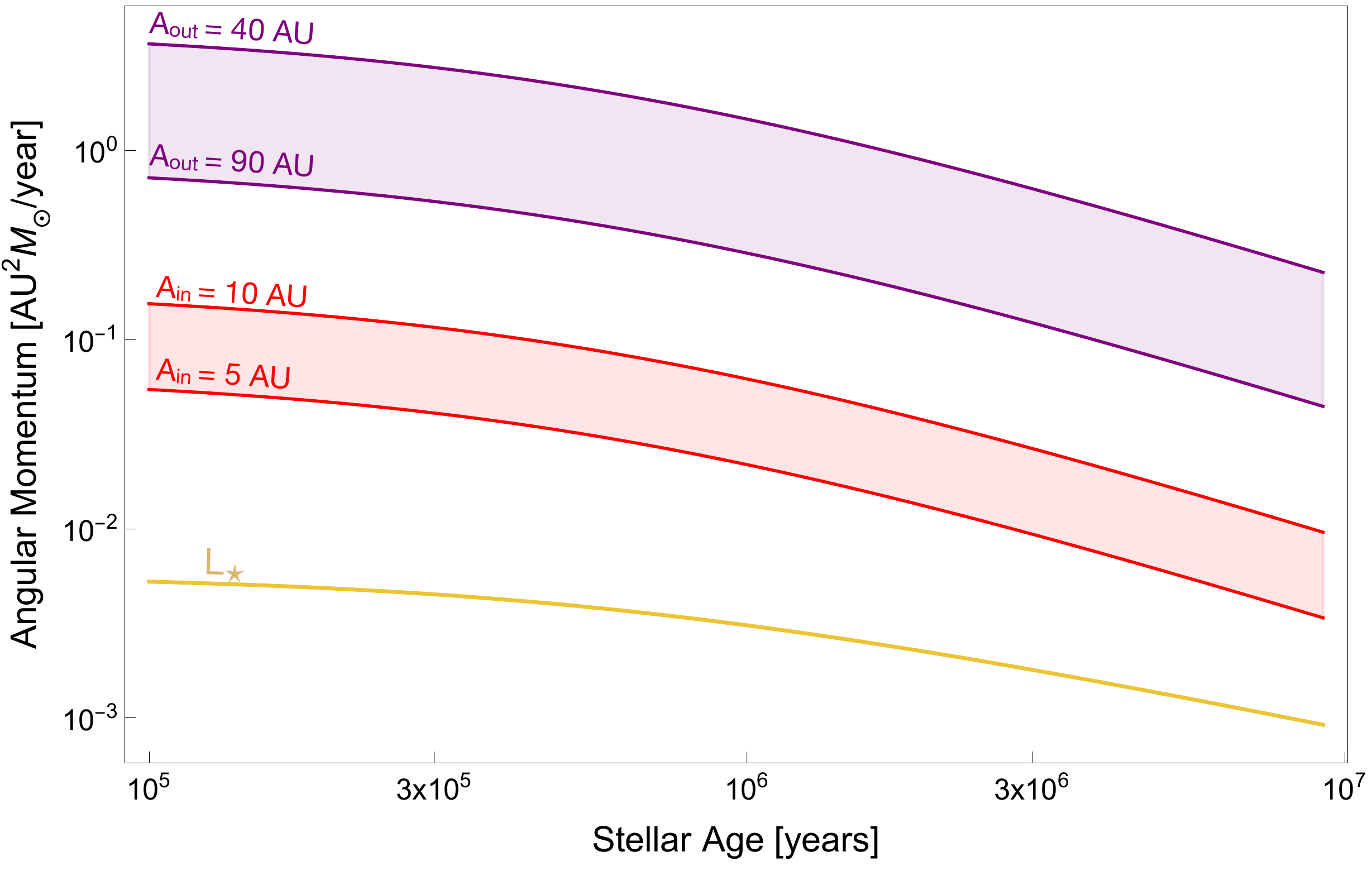}
 \caption{The angular momentum over time for three components: a range of possible values for the inner disk in purple, with an outer edge ranging from 5 to 10 AU; an outer disk, denoted in red, with inner edge varying between 40 and 90 AU; and a solar mass star in yellow. Within the time that the disk is expected to be present in the system, the hierarchy of angular momentum described in Section \ref{sec:l_hierarchy} is expected to hold for this typical architecture. }
    \label{fig:l_hierarchy}
\end{figure}

Throughout the evolution of the system shown in Figure \ref{fig:l_hierarchy}, there emerges a clear hierarchy in the angular momenta, which can thus be written as follows:

\begin{equation}
    \mathcal{L}_{\rm{d,out}} \gg \mathcal{L}_{\rm{d,in}} \gg \mathcal{L}_{*}  \label{eqn:hierarchy}\,, 
\end{equation}
where the difference between each angular momentum generally constitutes an order of magnitude. 

Satisfaction of Equation (\ref{eqn:hierarchy}) allows us to adopt a simplified mathematical relationship between the three components: the outer disk component, with the largest angular momentum, will essentially serve as a static potential on the inner two components. In turn, the inner disk will be torqued only by the outer disk component, and the star will be torqued in turn by the inner disk. It is important to note that violation of this assumption would force more complex dynamical behavior including back reactions of interior components on exterior components.

\section{Gravitational Torques}\label{sec:grav}

The angular momentum arguments of the previous section illustrate that the dynamics of the broken disk-star system are fundamentally analogous to the gravitational interaction between a star, intact disk and companion discussed elsewhere \citep{Batygin2012,2013ApJ...778..169B,Spalding&Batygin2014, Lai2018}. Nevertheless, a key distinction in the governing dynamics arises from the time-dependence of the outer disk's mass. As in the binary-disk model, we adopt a Hamiltonian framework in describing the system dynamics. Because this model is well developed in other works, we focus our attention on the characteristic frequencies, namely the precession rate of the inner disk and stellar axis, which are unique to the broken disk set up. For a more detailed description of the Hamiltonian geometry we refer the reader to the works cited above. 

\subsection{Disk Precession}\label{subsec:diskrate}

To begin, we consider the gravitational response of the inner disk to a misaligned outer disk. As in Section (\ref{sec:l_hierarchy}), we approach the inner and outer disks as a series of infinite annular wires whose dynamical interactions are limited to the idealized case where viscosity, bending waves, and self-gravity are sufficiently large that the disk remains coupled and coplanarity is maintained among neighboring annuli. 

Taking advantage of the smallness of the inner disk relative to the outer disk, we adopt a Kaula-type expansion of the Hamiltonian, using the semi-major axis ratio ($a_{\rm{in}}/a_{\rm{out}}$) as an expansion parameter. Note that an expansion of this type is consistent with the aforementioned angular momentum hierarchy and places no restriction on the mutual inclination of the inner and outer disks. To quadrupole order, the Hamiltonian is written \citep{Mardling2010, Kaula1962}:
\begin{equation}
\mathcal{H}_{\rm{wire}} = \frac{G}{4}\frac{dm_{\rm{in}} dm_{\rm{out}}}{a_{\rm{out}}}\left(\frac{a_{\rm{in}}}{a_{\rm{out}}} \right)^2 \left[ \frac{1}{2} \left(3 \cos^2\left(i_{\rm{in}}\right) -1 \right) \right]\,,
\label{eqn:hammy1.1}
\end{equation}
where $i_{\rm{in}}$ is the inclination of the inner disk and we have aligned our frame of reference with the plane of the outer disk such that $i_{\rm{out}} = 0$. 

Although the Keplerian orbital elements are widely understood and readily interpretable, they do not constitute a canonically conjugated set of coordinates. In order to solve for the evolution of the system by way of Hamilton's equations, we must introduce an alternate, canonical coordinate system, namely the scaled Poincaré action-angle coordinates: 
\begin{equation}
Z_{\rm{in}} = 1 - \cos{\left( i_{\rm{in}} \right)} \,,
\qquad 
z_{\rm{in}} = -\Omega_{\rm{in}}\,.
\label{eq:poincare}
\end{equation}
Because these are not the standard Poincaré coordinates, but are scaled by the angular momentum of an annulus ($ d \Lambda_{\rm{in}} = {dm_{\rm{in}} \sqrt{G M a_{\rm{in}}}}$), for the Hamiltonian to remain canonical it must be similarly scaled. After substituting coordinates and scaling, the new scaled Hamiltonian becomes: 
\begin{equation}
\begin{split}
\Tilde{\mathcal{H}}_{\rm{wire}} &= \frac{\mathcal{H}_{\rm{wire}}}{d\Lambda_{\rm{in}}} \\&=  \frac{3 }{4}\sqrt{\frac{G M}{a_{\rm{in}}^3}} \frac{dm_{\rm{out}}}{M}\left(\frac{a_{\rm{in}}}{a_{\rm{out}}}\right)^3 \left(Z_{\rm{in}} - \frac{Z_{\rm{in}}^2}{2}\right)\,.
\label{eqn:hammy1.2}
\end{split}
\end{equation}
 Note, however, that Hamiltonian (\ref{eqn:hammy1.2}) models only the gravitational forcing due to an outer disk annulus, and fails to account for the radial extent of the outer disk. The potential of the full outer disk is modeled by integrating the Hamiltonian radially from the outer-most edge of the disk to the outer edge of the disk gap ($A_{\rm{out}}$), 
\begin{equation}
\begin{aligned}
\begin{split}
    \mathcal{\Tilde{H}} &= \int_{A_{\rm{out}}}^{\infty} \mathcal{\Tilde{H}}_{\rm{wire}} \\
    &= \frac{3 \pi}{4} \sqrt{\frac{G M}{a_{\rm{in}}^3}} \frac{\Sigma_0 a_0 a_{\rm{in}}}{M} \left(\frac{ a_{\rm{in}}}{A_{\rm{out}}}\right)^2 \left(Z_{\rm{in}} - \frac{Z_{\rm{in}}^2}{2}\right)\,.
    \label{eqn:hammy1.3}
    \end{split}
    \end{aligned}
\end{equation}
Where we have used the stiff dependence of Equation (\ref{eqn:hammy1.2}) on $a_{\rm{out}}$, integrating the outer disk radially from infinity so that we can define the outer disk by the outer gap edge ($A_{\rm{out}}$) only. For a disk extending to $\sim 100$ AU this amounts to a correction factor of $< 1 \%$ and can be neglected for the purposes of this work.

Critically, the Hamiltonian in Equation (\ref{eqn:hammy1.3}) has no $z_{\rm{in}}$ dependence. From Hamilton's equations we understand this to mean that the inclination of the inner disk is constant in time and the longitude of ascending node of an inner disk annulus regresses at a constant rate, $dz_{\rm{in}}/dt$, calculated as
\begin{equation}
\begin{split}
\frac{d z_{\rm{in}}}{d t} = \frac{\partial \Tilde{\mathcal{H}}}{\partial Z_{\rm{in}}} = \frac{3 \pi}{4} \sqrt{\frac{G M}{a_{\rm{in}}^3}} \frac{\Sigma_0 a_0 a_{\rm{in}}}{M} \left(\frac{ a_{\rm{in}}}{A_{\rm{out}}}\right)^2 \left(1 - Z_{\rm{in}}\right)\,.
\label{eqn:dzdt}
\end{split}
\end{equation}
To find the rate of nodal recession for the \textit{entire} inner disk, we approximate the mass, 
\begin{equation}
\begin{split}
m_{\rm{in}} = \int_{a_{\rm{x}}}^{A_{\rm{in}}}{2 \pi \Sigma_0 a_0  da_{\rm{in}}} \approx 2 \pi \Sigma_0 a_0 A_{\rm{in}}\,,
\end{split}
\end{equation}
under the assumption that the inner disk has significant radial extent ($a_\text{X } \ll A_{\text{in}}$). The rate of nodal precession may be derived by taking the orbital angular momentum weighted average of the forced recession rates of the individual disk annuli
\begin{equation}
\begin{split}
\omega =& \frac{\int_{a_{\rm{X}}}^{A_{\rm{in}}} \frac{d z_{\rm{in}}}{d t} 2 \pi \Sigma_0 a_0 \sqrt{G M a_{\rm{in}}} da_{\rm{in}}}{ \int_{a_{\rm{X}}}^{A_{\rm{in}}}2 \pi \Sigma_0 a_0 \sqrt{G M a_{\rm{in}}} da_{\rm{in}}}
\\&= -\frac{3}{16} \sqrt{\frac{G M}{A_{\rm{in}}^3}} \frac{m_{\rm{in}}}{M} \left(\frac{ A_{\rm{in}}}{A_{\rm{out}}}\right)^2 \cos{(i_{\rm{in}})}\,.
\label{eqn:omega}
\end{split}
\end{equation}
The result is that the nodal precession of the entire disk ($\omega$) differs from an annulus at the outer edge of the inner disk by a factor of $1/2$.

Although our approach is distinct, Equation (\ref{eqn:omega}) is analogous to Equation (19) in  \citet{1996MNRAS.282..597L}, where the precession frequency was derived by dividing the total torque acting on the inner disk by the disk's angular momentum. Notably, by assuming a consistent density profile between the inner and outer disks, Equation (\ref{eqn:omega}) does not explicitly include the mass of the outer disk forcing the inner disk. Despite this somewhat less intuitive formulation, Equation (\ref{eqn:omega}) has the advantage of describing the dynamics of the inner disk forced by the outer disk by the boundaries of the disk gap \textit{only}.

Bear in mind, Equation (\ref{eqn:omega}) applies only if the inner disk precesses as a rigid body. That is, inclination ($i_{\rm{in}}$) and node ($\Omega_{\rm{in}}$) are consistent between disk annuli. This assumption is appropriate so long as the disk precession timescale exceeds the time for sound waves to propagate through the disk, or $(c_s(r)/r)>\omega$, where $c_s(r)$ is the sound speed in the disk at radius $r$ \citep{1996MNRAS.282..597L}. For a thin, flat disk $c_s\propto r^{-3/8}$ \citep{Armitage2020}, and the lower bound $c_s(r = 10 \rm{AU})/10\rm{AU} \approx 4 \times 10^{-10}$ is over three orders of magnitude larger than the largest precession frequency ($\omega$) relevant here. Thus, we can safely assume the disk precesses as a rigid body and proceed with our analysis. 

\subsection{Stellar Spin Axis Precession}

For the sake of simplicity, we model the rotational deformation of the central star as a point mass surrounded by a circular wire. To quadrupole order, these two systems can be considered mathematically equivalent so long as their moments of inertia and gravitational potentials are consistent. The quadrupolar potential of a rotationally deformed spheroid is
\begin{equation}
  V_{\text{quad}} = \frac{G M}{r}J_2\left(\frac{R}{r}\right)^2 \mathcal{P}_2\left(\cos(\theta)\right)\,,
    \label{eq:quad}
\end{equation}
where $\mathcal{P}_2\left(\cos(\theta)\right)$ is a Legendre polynomial of degree 2 and the angle $\theta$ is measured from the stellar spin axis (Section 4.5 of \citet{murray1999solar}).  $J_2$ is a dimensionless constant used here in its approximate form 
\begin{equation}
    J_2 = \frac{\Omega^2 R^3}{3 G M} k_2\,,
\end{equation}
where $k_2$ is the stellar Love number. Setting the potential of the hoop to Equation (\ref{eq:quad}) and the moment of inertia to $I_0 M R^2$, the appropriate semi-major axis of the hoop may be algebraically determined,
\begin{equation}
a_{\rm{h}} = \left( \frac{16 \Omega^2 k_2^2 R^6}{9 I_0^2 G M}\right)^{1/3}\,.
\label{eq:ahoop}
\end{equation}
 Note that Equation (\ref{eq:ahoop}) is identical to Equation (12) provided in the Supplementary Information of \citet{Batygin2012}. However, unlike \citet{Batygin2012}, we assume a fully convective T Tauri star, with a polytropic index of $n = 3/2$. For such a star the Love number and moment of inertia factor have characteristic values $I_0 = 0.21$ and $k_2 = 0.14$ \citep{Chandrasekhar1939}. 

By modelling the oblate central star as a circular wire forced by an external disk, the problem with which we are concerned here is nearly identical to that described in the previous section. Our analysis can thus proceed in much the same way, with the caveat that the inclination of the external perturber, in this case the inner disk, is no longer set to zero but maintains some constant value throughout its evolution. In orbital elements, the Hamiltonian is written as \citep[Equation A3 of][]{Mardling2010}
\begin{equation}
\begin{split}
\mathcal{K}_{\rm{wire}} =& - \frac{1}{4}\frac{G dm_{\rm{in}} m_{\rm{h}}}{a_{\rm{in}}} \left(\frac{a_{\rm{h}}}{a_{\rm{in}}}\right)^2\\& \times \bigg[\frac{1}{4} \left(3 \cos^2\left(i_{\rm{in}}\right) -1\right)\left(3 \cos^2\left(i_{\rm{h}}\right) -1\right) \\&+ \frac{3}{4} \sin{\left(2 i_{\rm{in}}\right)}\sin{\left(2 i_{\rm{h}}\right)}\cos{\left(\Omega_{\rm{in}}-\Omega_{\rm{h}}\right)} \\&+ \frac{3}{4} \sin^2\left(i_{\rm{in}}\right) \sin^2\left(i_{\rm{h}}\right) \cos \left(2 \left(\Omega_{\rm{in}} - \Omega_{\rm{h}}\right)\right) \bigg].
\end{split}
\label{eqn:hammy2.1}
\end{equation}
Again, we exchange orbital elements for Poincaré coordinates, integrate radially with respect to the inner disk ($a_{\rm{in}}$) and scale Equation (\ref{eqn:hammy2.1}) by the angular momentum of the stellar hoop $\Lambda_{\rm{h}} = m_{\rm{h}} \sqrt{G M a_{\rm{h}}}$, to find the Hamiltonian: $\Tilde{\mathcal{K}}_{\rm{wire}} = \mathcal{K}_{\rm{wire}}/\Lambda_{\rm{h}}$. 
This form of the Hamiltonian given in \citet{Mardling2010} assumes circular orbits for the disk components.
The final pre-factor of the Hamiltonian gives the characteristic nodal precession frequency of the stellar spin axis, which we calculate to be, 

\begin{equation} 
f = \frac{3 \pi}{4}\sqrt{\frac{G M}{a_{\rm{h}}^3}}\frac{\Sigma_0 a_0 a_{\rm{x}}}{M}\left(\frac{a_{\rm{h}}}{a_{\rm{x}}}\right)^3\,.
\label{eqn:f}
\end{equation}

As mentioned in Section (\ref{subsec:diskrate}), Equation (\ref{eqn:hammy1.3}) is independent of $z_{\rm{in}}$. This means the inner disk inclination with respect to the binary plane, ($i_{\rm{in}}$), is conserved, while the inner disk node precesses at a constant rate ($\omega$). The Hamiltonian ($\Tilde{\mathcal{K}}$) is therefore a non-autonomous system with a particularly simple time dependence ($z_{\rm{in}} = \omega t$). The Hamiltonian can be made autonomous with a correspondingly straightforward canonical transformation into a frame precessing with the node of the inner disk. To boost into this new frame, we employ a generating function of the second kind \citep{Morbidelli2002}:
\begin{equation} 
G_2 = \left(z_{\rm{h}} - \omega t\right)\Phi,
\end{equation}
where $\phi = \left(z_{\rm{h}} - \omega t\right)$ is the canonical coordinate and the canonical momentum is determined using the relation
\begin{equation}
Z_{\rm{h}} = \frac{\partial G_2}{\partial z_{\rm{h}}}= \Phi.
\end{equation}
The Hamiltonian is then transformed according to: 
\begin{equation}
    \bar{\mathcal{K}} = \Tilde{\mathcal{K}} + \frac{\partial G_2}{\partial t}.
\end{equation}
Having removed explicit time dependence we also scale the Hamiltonian by $\omega$ so that all evolving parameters are contained in a single pre-factor $f/\omega$ and the Hamiltonian takes the form 
\begin{equation} 
\begin{split}
\bar{\mathcal{K}}=&-\Phi+\frac{f}{\omega} \bigg[ \frac{3}{4}\left(\frac{2}{3} - 2 Z_{0}+Z_{0}^2 \right)\left(\frac{2}{3} - 2 \Phi+\Phi^2 \right)
\\&+ \left( Z_{0} - 1\right)\sqrt{Z_{0}\left(2 - Z_{ 0}\right)} \left( \Phi - 1\right)
\sqrt{\Phi\left(2 - \Phi\right)}\\&\times \cos{\left(\phi\right)} + \frac{1}{4}Z_0 \left(Z_0 -2\right)\Phi\left(\Phi -2\right)\cos\left(2\phi\right)\bigg]\,.
\end{split}
\label{eqn:hammy2.2}
\end{equation}
Invoking Hamilton's equations, the evolution of the stellar spin axis is governed by the partial derivatives of Equation (\ref{eqn:hammy2.2}) with respect to canonical coordinates $\Phi$ and $\phi$. And the stellar obliquity is defined by the dot product of the stellar angular momentum and disk angular momentum vectors. 

\begin{figure}
\centering
 \includegraphics[width=3.35 in]{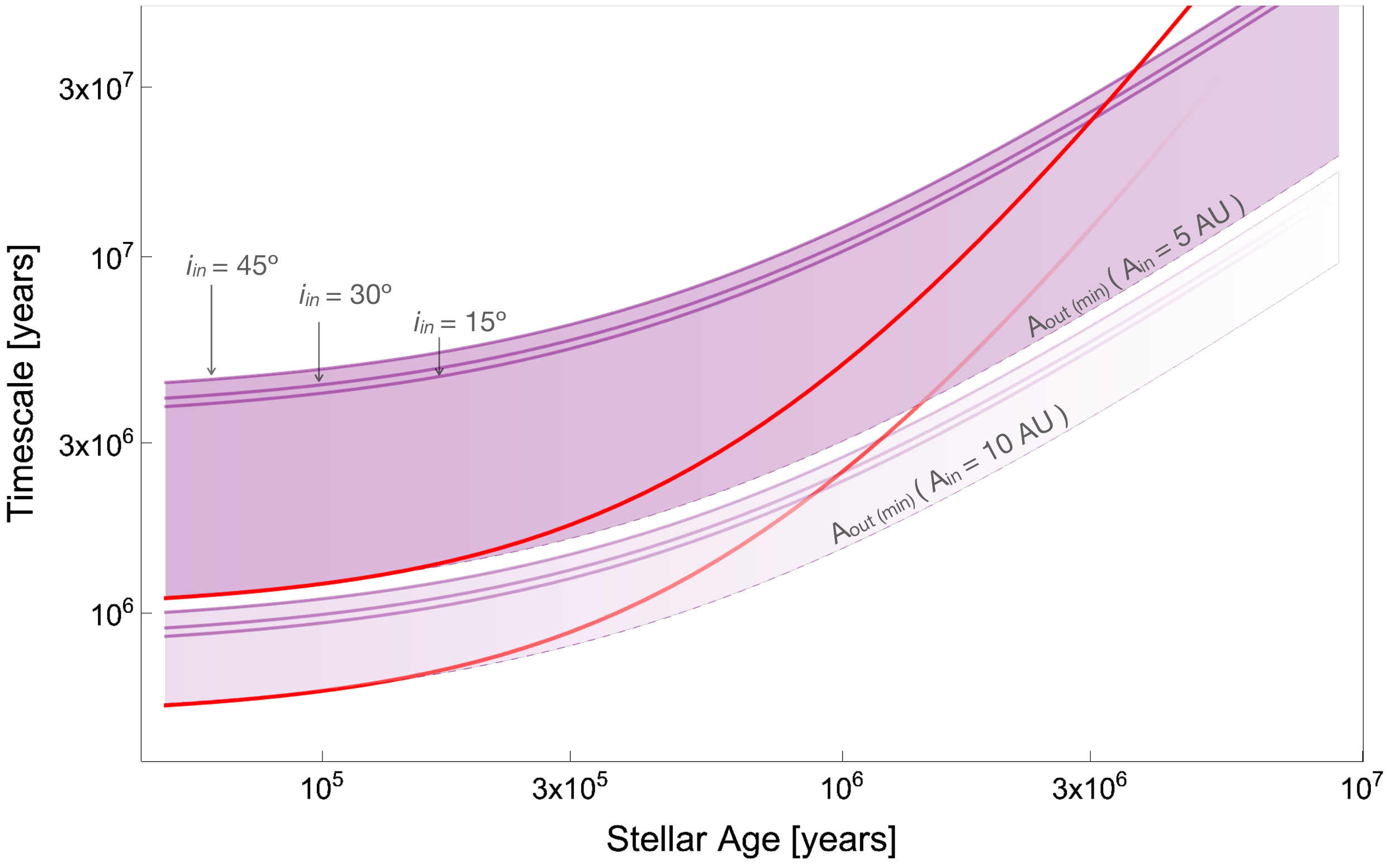}
 \caption{Evolution of timescales associated with the stellar spin axis ($2 \pi/f$) (red) and the inner disk ($2 \pi / \omega$)(purple). Multiple disk geometries are illustrated. An opacity gradient is applied for an inner gap edge $A_{\rm{in}} = 10$ AU, for which the time to resonant crossing of $f$ and $\omega$ is shorter. The shaded purple regions describe inner disk precession for outer gap ($A_{\rm{out}}$) ranging from 80 AU to $A_{\rm{out(min)}}$ determined by Equation (\ref{eqn:Aoutmin}). Solid purple contours identify the boundaries of ($2 \pi/\omega$) for $i_{\rm{disk}} = 15^\circ\text{, } 30^\circ\text{, and }45^\circ$.  }
    \label{fig:timescales}
\end{figure}

\section{Obliquity Excitation}
\label{sec:obliq}

The Hamiltonian in Equation (\ref{eqn:hammy2.2}) is characterized by two physical timescales, the inner disk precession timescale ($2 \pi/ \omega$) and the stellar spin axis precession time scale ($2 \pi /f$). Obliquity excitation is expected to occur when $f$ and $\omega$ evolve such that their ratio crosses unity from above, as $f>\omega$ at early times.  Using Equations (\ref{eqn:omega}) and (\ref{eqn:f}), the ratio ($f/\omega$)  simplifies to: 

\begin{equation}
    \frac{f}{\omega} = \frac{8}{3 \cos{\left(i_{\rm{in}}\right)}}  \frac{k_2}{I}\left(\frac{R_\star}{a_x}\right)^3 \left(\frac{a_x}{A_{in}}\right)^{3/2}\left(\frac{A_{out}}{a_x}\right)^2\,.
\end{equation}
Where we have substituted Equation (\ref{eq:ahoop}) for $a_{\rm{h}}$ and assumed that the star is co-rotating with the inner edge of the inner disk, or $\Omega = \sqrt{G M/ a_{\rm{x}}^3}$. In the case of this second substitution, we have imposed a ``disk-locked'' condition, in which the disk truncation radius is coincident with the radius at which the relative velocity of the stellar magnetosphere and disk material is zero \citep{1991ApJ...370L..39K, Shu1994, Mohanty&Shu2008}. While the disk-locked is an idealization that ignores the complexities of magnetic disk-star coupling, it is to within an order of magnitude correct, and is therefore an appropriate simplification in the context of this work. 

The evolution of the characteristic timescales ($2 \pi/f$ and $2 \pi/\omega$) over the lifetime of the disk are shown in Figure (\ref{fig:timescales}). From the figure it is clear that as the inner and outer disk dissipate both timescales increase, but the stellar precession timescale ($2 \pi/ f$) is additionally dependent on stellar contraction. That is, $f$ decreases more steeply with time than $\omega$ and crossing is inevitable so long as $f > \omega$ in the beginning of the evolution. By setting $f/\omega = 1$ and $t = 0$, we  can solve for the minimum value of the outer edge of the disk gap ($A_{\rm{out\left(min\right)}}$) for which obliquity excitation can be expected to occur,
\begin{equation}
\begin{split}
    A_{\rm{out(min)}} &\approx 33 \text{ AU } \sqrt{\cos{\left(i_{\rm{in}}\right)}}  \left[\frac{A_{\rm{in}}}{\text{AU}}\right]^{1/4} \left[\frac{M_\star}{M_\sun}\right]^{-1/4} 
    \\& \times \left[\frac{a_0}{\text{AU}}\frac{\Sigma_0}{2 \times 10^4 \text{ g}/\text{cm}^3}\right]^{-1/2}\left[\frac{R_0}{4 R_\sun}\right]^{3/2} \\& \times \left[\frac{B_{\rm{s}}}{.1 \text{ T}}\right]\left[\frac{\tau_{\rm{acc}}}{.5 \text{ Myr}}\right]^{1/2}\,.
\end{split}
\label{eqn:Aoutmin}
\end{equation}
Note that $A_{\rm{out(min)}}$ is consistent with the angular momentum hierarchy outlined in Section (\ref{sec:l_hierarchy}), remaining above 40 AU for $i_{\rm{in}}\leq50^\circ$ and $i_{\rm{in}}\leq63^\circ$ when $A_{\rm{in}} = 5 \text{ AU}$ and $10 \text{ AU}$ respectively. Additionally, as Figure (\ref{fig:timescales}) illustrates, for all relevant disk parameters $f/\omega$ crosses unity well within a maximum disk lifetime of $\sim 10^7$ years. 

\begin{figure}

 \includegraphics[width=3.35 in]{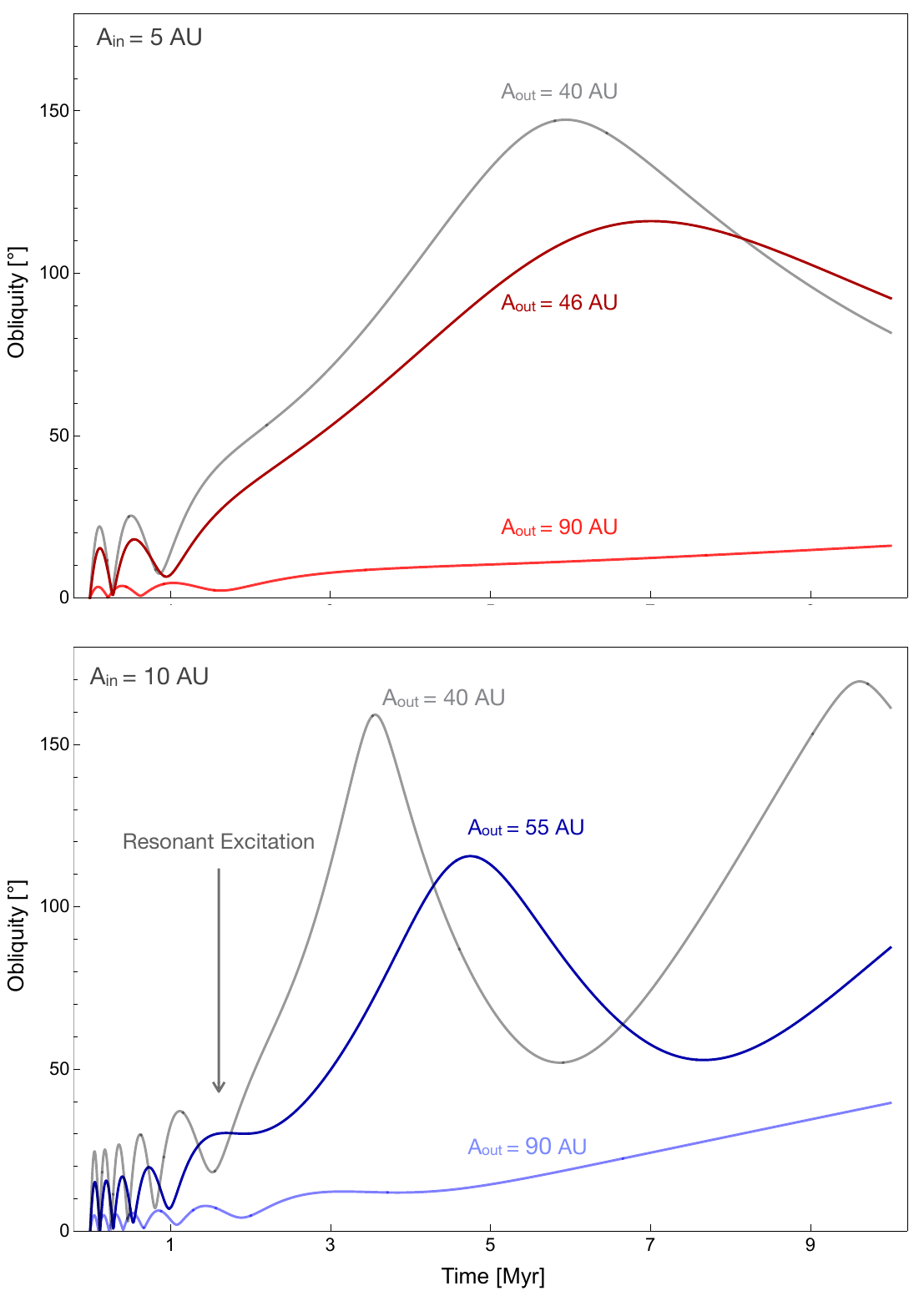}
 \caption{Obliquity evolution of stellar spin axis was calculated by numerically solving Hamilton's equations with respect to the canonical variables $\Phi$ and $\phi$ and the Hamiltonian in Equation (\ref{eqn:hammy2.2}). The evolution of several different disk geometries is shown -- in red and blue $A_{\rm{in}} = 5$ AU and 10 AU respectively. Integrations plotted in blue and red have $i_{\rm{disk}} = 30^\circ$. Lighter tones differentiate integrations with $A_{\rm{out}} = 90$ AU and in darker toned integrations  $A_{\rm{out}} = A_{\rm{out(min)}}$. Gray curves indicate $A_{\rm{out}}$ is set to 40 AU and $i_{\rm{disk}}$ set by Equation (\ref{eqn:Aoutmin}) to $50^\circ$ and $63 ^\circ$ in the upper and lower panel respectively.}
\label{fig:numerical}
\end{figure}

Having demonstrated that the relevant timescales are consistent with obliquity excitation, we can use the analytical framework developed in Section (\ref{sec:grav}) to explore the outcome of the encounter for different system configurations. Six example evolutions are shown in Figure (\ref{fig:numerical}). The variable parameters compared include the disk gap bounds ($A_{\rm{in}}$ and $A_{\rm{out}}$) and the inner disk inclination ($i_{\rm{in}}$). In the upper panel, evolved systems have $A_{\rm{in}}$ set to 5 AU, and in the lower panel $A_{\rm{in}} = 10$ AU. In both panels, curves in dark tones indicate a disk inclination of $30^\circ$ with $A_{\rm{out}}$ minimized according to Equation (\ref{eqn:Aoutmin}). Evolutions plotted in gray have $A_{\rm{out}}$ set to 40 AU and disk inclination set according to Equation (\ref{eqn:Aoutmin}). For systems plotted lighter tones, $A_{\rm{out}}$ is maximized according to angular momentum constraints and set to 90 AU. 

The initial dynamical behavior is consistent between all the systems considered in Figure (\ref{fig:numerical}): early in the evolution efficient angular momentum transport between the inner disk and star leads to gravitational coupling. The stellar spin axis trails just behind the precessing angular momentum vector of the inner disk, resulting in small oscillations in obliquity. 

As the systems evolve, and the ratio of disk and stellar precession frequencies ($f/\omega$) approaches unity, they experience a distinct jump in obliquity. The amplitude of this excitation ranges from tens of degrees to nearly $180^\circ$, where the primary factor distinguishing a large jump in obliquity from a small jump is the time of resonant encounter. By decreasing the disk gap width, we decrease the precession timescale ratio ($f/\omega$) and the time to resonant crossing. When the excitation occurs late in the evolution, the star is significantly less oblique and the inner and outer disks have largely dissipated -- resulting in significantly diminished gravitational torques and muted obliquity excitation. And while extreme obliquities can be achieved via this disk torquing mechanism, the parameters required press the resonant crossing boundary laid out in Equation (\ref{eqn:Aoutmin}). That is, while obliquities can clearly occur as a result of broken disk torquing, the mechanism is fundamentally less severe than the obliquities produced in the analogous binary torquing case by virtue of their occurring later in the disk lifetime. 

\section{Conclusion}
\label{sec:conclude}

In this work, we have identified a pathway by which obliquities may be excited in the context of a pure disk-star system. Building from the binary-torquing framework \citep{ Batygin2012, 2013ApJ...778..169B, Spalding&Batygin2014, Lai2018}, we analyze stellar spin axis forcing in the potential of a broken disk, and show that large spin-orbit misalignments can be generated without invoking a binary companion. For nominal parameters, well within the observational range described by \citet{Ansdell2020, Francis2020}, we show that conditions for resonant excitation are generally met. 

While our results demonstrate that a misaligned, broken disk is capable of torquing the host star out of alignment, in developing our model we have employed several simplifications in an effort to enhance the clarity of our model. In our estimation, these complicating factors will not change the overall outcome, but they are critical to note. In particular, we have assumed here that the boundary of the disk gap remains static throughout the lifetime of the disk. The origin and evolution of disk gaps remains an active area of investigation, but the radial structure is likely in flux early on in the disk lifetime \citep{vanderMarel2018, vanderMarel2019}. The relatively late onset of resonant excitation seen in the disk-disk torquing scenario suggests that early variation in the radial gap structure are unlikely to have a significant effect on the excitation of stellar obliquity. 

Additionally, the Hamiltonian given in Equation (\ref{eqn:hammy2.2}) neglects non-conservative effects in the disk-disk system. Averaging of gravitational torques over the precession timescale of the stellar spin axis create a dynamical equilibrium aligned with the angular momentum vector of the outer disk. Dissipative processes such as accretion and Magnetic torquing, force the spin axis closer to this equilibrium \citep{2011morbi, 2015ApJ...811...82S}. Interestingly, this means that even if the effects of dissipation overwhelmed the resonant excitation, the stellar spin axis would be expected to be misaligned with respect to the inner disk. 

The object of this work has been to establish broken disk torquing as a viable alternative pathway to stellar obliquities. We note, however, that misalignments generated via disk torquing may differ fundamentally from their binary generated counterparts. Because the inner and outer disks evolve along a shared trajectory, disk dissipation has significantly less impact on the overall system evolution than stellar contraction. The evolutionary track, and by extension stellar type, of the star in question thus retains an outsized role in obliquity generation. The extended contraction timescale for an M dwarf, for instance, limits the resonant crossing time to the end stages of the disk lifetime at which point the diminished gravitational torques do not permit large obliquity excitation. 

In the case of T Tauri stars explored in this work, the evolutionary track does provide opportunity for the generation of misalignments. However, the singular dependence on the contraction timescale means that the amplitude of obliquity excitations are largely dependent on disk gap parameters and require more tuning to achieve high amplitudes than those produced by the binary torquing mechanism. A detailed study of the ways in which the results of these two processes diverge is outside of the scope of this work but may provide avenues for future study.

\medskip
\textbf{Acknowledgements.} We would like to thank our referee, Rosemary Mardling, whose thoughtful insights significantly improved this work. J.C.B.~has been supported by the Heising-Simons \textit{51 Pegasi b} postdoctoral fellowship. M.E.M.~has been supported by the \textit{Graduate Fellowship for STEM Diversity}. This research is based in part upon work supported by NSF grant AST 2109276. K.B. thanks the David and Lucile Packard Foundation for their generous support. 

Software: pandas \citep{ mckinney-proc-scipy-2010}, IPython \citep{PER-GRA:2007}, matplotlib \citep{Hunter:2007}, scipy \citep{scipy}, numpy \citep{oliphant-2006-guide}, Jupyter \citep{Kluyver:2016aa}, Mathematica \citep{mathematica}

\bibliographystyle{mnras}
\bibliography{refs}

\end{document}